\documentclass[twocolumn,showpacs,showkeys,superscriptaddress,preprintnumbers,amsmath,amssymb,prl]{revtex4}
\usepackage{graphicx}
\usepackage{dcolumn}
\usepackage{bm}
\usepackage{amsmath}
\begin{document}

\title{Particle Scale Dynamics in Granular Impact}

\author{Abram H. Clark}
\affiliation{Department of Physics and Center for Nonlinear and Complex Systems, Duke University, Durham, North Carolina 27708, USA}
\author{Lou Kondic}
\affiliation{ Department of Mathematical Sciences, New Jersey Institute of Technology, Newark, New Jersey 07102, USA}
\author{Robert P. Behringer}
\affiliation{Department of Physics \& Center for Nonlinear and Complex Systems, Duke University, Durham, NC 27708, USA}

\begin{abstract}
We perform an experimental study of granular impact, where intruders strike 2D beds of photoelastic disks from above. High-speed video captures the intruder dynamics and the local granular force response, allowing investigation of grain-scale mechanisms in this process. We observe rich acoustic behavior at the leading edge of the intruder, strongly fluctuating in space and time, and we show that this acoustic activity controls the intruder deceleration, including large force fluctuations at short time scales. The average intruder dynamics match previous studies using empirical force laws, suggesting a new microscopic picture, where acoustic energy is carried away and dissipated.
\end{abstract}
\date{\today}


%


\maketitle
The penetration of a dense granular material by a high-speed intruder
occurs routinely in meteor and ballistic impacts. Many previous
studies \cite{Euler1745,Poncelet1829,Allen1957,Forrestal1992,Tsimring2005,Katsuragi2007,Goldman2008,Goldman2010}, both recent and dating back to Euler and Poncelet, have used variations of a macroscopic
force law:
\begin{equation}
F=m\ddot{z}=mg-f(z)-h(z)\dot{z}^2.
\label{eqn:forcelaw}
\end{equation}
Here, $z$ is the intruder depth relative to the top of the original,
unperturbed surface (i.e., $z=0$ at initial impact), $mg$ is the
gravity force, $f(z)$ characterizes hydrostatic effects, $h(z)$
is often assumed constant, $h(z) = b$, and dots denote time derivatives. In
Eq.~(\ref{eqn:forcelaw}), $h(z) \dot{z}^2$ represents a coarse-grained collisional stress. We note that other effects, including a
depth-dependent Coulomb friction term have been proposed
\cite{Tsimring2005,Katsuragi2007}. Despite the success of
extensive previous studies \cite{Euler1745,Poncelet1829,Allen1957,Forrestal1992,Tsimring2005,Katsuragi2007,Goldman2008,Goldman2010,Ciamarra2004,Ambroso2005,Bruyn2004,Walsh2003,Nelson2008,Newhall2003,Hou2005}, the connections between the local granular
response, the microscopic processes responsible for dissipating
kinetic energy, and the dynamics of the intruder are still subjects of
debate, largely due to experimental difficulties in obtaining
sufficiently fast data at small scales.

\begin{figure*}[th!] 
 \includegraphics[clip,trim=10mm 5mm 10mm 3mm,width=\textwidth]{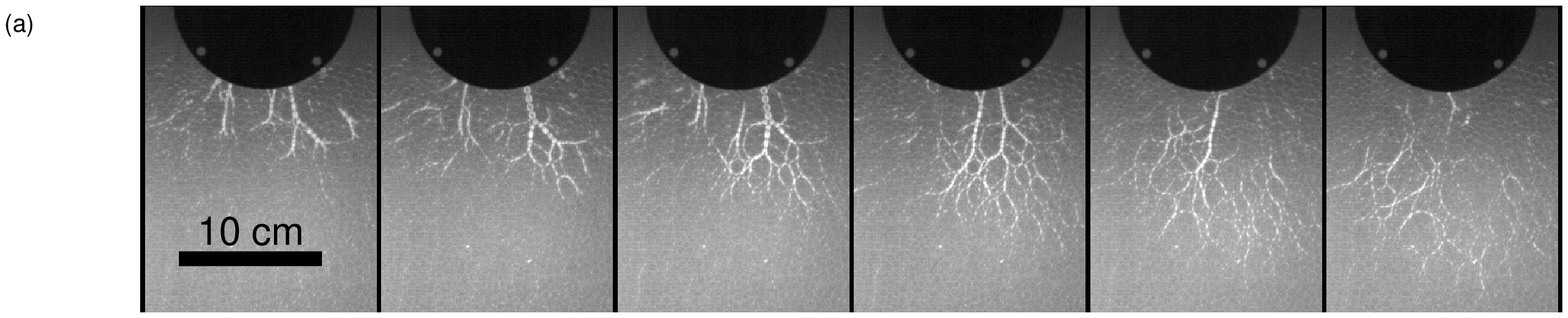}
 \includegraphics[clip,trim=10mm 5mm 10mm 0mm,width=\textwidth]{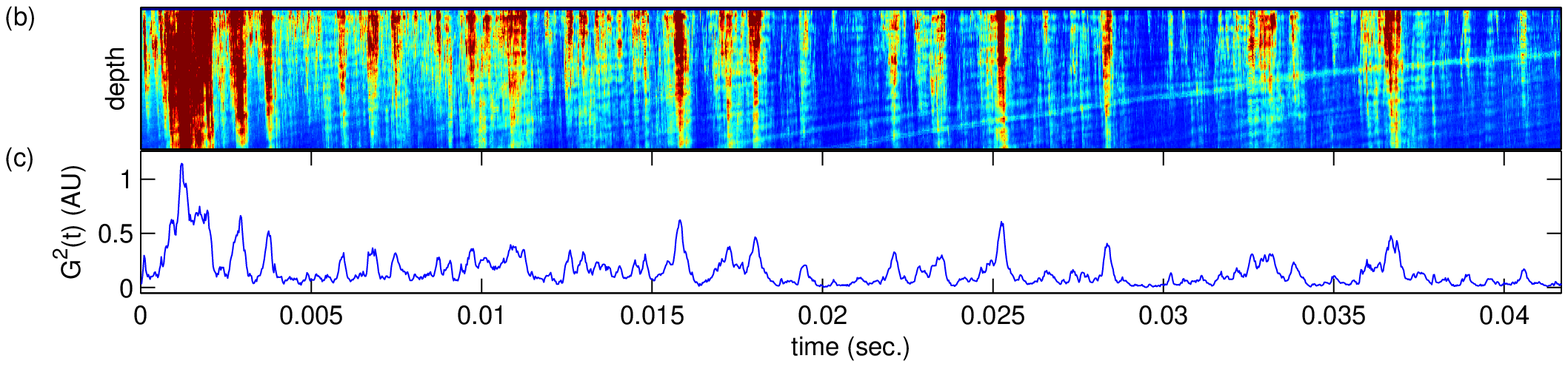}
 \caption{(color online). (a) Six selected frames, starting at
     2.75 ms after impact and spanning 475 $\mu$s, showing the 
     end of a typical compressional event which generates an
     acoustic pulse, which disconnects from the intruder. (b) A space-time plot 
     of $G^2$ in an angular region under the bottom half of
     the intruder (half-annulus) over time. The \emph{x} axis is time, and
     the \emph{y} axis is radial distance from the bottom of the intruder,
     where the top of the plot corresponds to the bottom edge of the
     intruder. The slope of the disturbances gives a consistent acoustic speed of $\sim 325$~m/s.
     (c) The sum of the response in the space-time plot above,
     after subtracting background inhomogeneities. Calibrating
     this will yield our measurement of instantaneous
     force, as shown later, where the range shown above [0 to 1.1
     arbitrary units (AU)] maps to an intruder acceleration range of 0 to 27
     g.}
 \label{fig:frames}
\end{figure*}

In this Letter, we address this issue experimentally by high-speed
imaging of an intruder of mass, $m$, which impacts a
quasi-two-dimensional system of photoelastic particles (bidisperse,
larger particle diameter $d$) at speeds $v_0 \leq 6.5$~m/s, yielding
both the intruder dynamics and the force response of individual
grains (Fig.~\ref{fig:frames}). Here, as in many previous experiments, $v\ll C$, where $C
\simeq 300$~m/s is the granular sound speed, measured 
from photoelastic space-time plots, as in Fig.~\ref{fig:frames}(b). The frame rates of $\sim C/d$ 
capture the microscopic granular response. The
primary intruder energy loss mechanism in these experiments is due
to intense, intermittent acoustic pulses traveling at speeds $\sim C$
along networks of grains, transmitting energy from the intruder into
the medium. These pulses decay roughly exponentially with distance
from the intruder. The force on the intruder is strongly fluctuating,
due to the intermittency of the force network or acoustic activity, but
the mean behavior is consistent with empirical models
used previously \cite{Euler1745,Poncelet1829,Allen1957,Forrestal1992,Tsimring2005,Katsuragi2007,Goldman2008,Goldman2010}.

\begin{figure}
	\includegraphics[clip,trim=7mm 0mm 10mm 7mm,width=0.85\columnwidth]{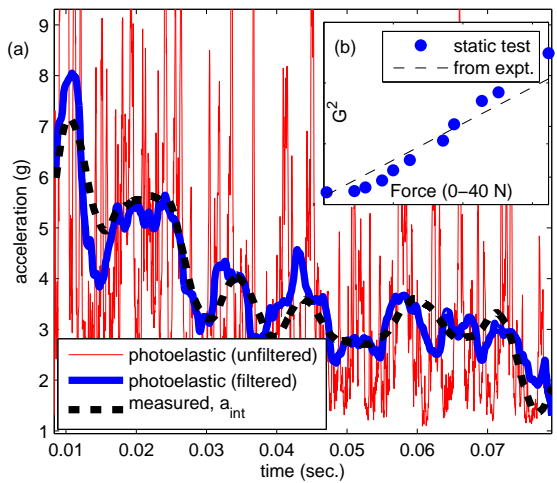}
	\caption{(color online.) (a) Comparing the intruder trajectory
            to the photoelastic response from Fig.~\ref{fig:frames}(c) shows that the intruder acceleration
            is very well correlated to the photoelastic or acoustic fluctuations in
            high-speed videos. We time-average the photoelastic response (thick, blue line)
            to match the time scale of the acceleration measurement $a_{int}$ (black,
            dashed line), which has limited time resolution. Rescaling the
            photoelastic measurement gives extremely
            close agreement with the measured deceleration (both the
            mean and fluctuations). The calibrated photoelastic force measurement without
            time filtering (thin, red line) shows much larger fluctuations at a
            much shorter time scale. (b) The inset shows the calibration of photoelastic
            response versus 2D pressure (force per width of intruder or piston) from experiment 
            (black dashed line) and from a static test (blue circles), with good agreement.}
	\label{fig:accphoto}
\end{figure}

{\em Experimental techniques.\textemdash}The experimental apparatus consists of
two thick Plexiglas sheets (0.91~m $\times$ 1.22~m $\times$ 1.25~cm),
separated by a thin gap (3.3~mm) which is filled by photoelastic
disks (thickness of 3~mm) of two different diameters (6 and 4.3
mm). These disks are cut from PS-1 material (Vishay Precision Group;
bulk density of 1.28~g/cm$^3$, elastic modulus of 2.5 GPa, and Poisson's
ratio of 0.38). Intruders are machined from a bronze sheet (bulk density
of 8.91~g/cm$^3$ and thickness of 0.23~cm) into disks of diameters $D$
of 6.35, 10.16, 12.7, and  20.32 (data for $D = 12.7$~cm intruder used in images and
time-series data shown here are typical for all $D$). These intruders
are dropped from a height $H \leq $~2.2~m, through a shaft
connected to the top of the thin gap containing the particles,
producing an impact speed $v_0 \simeq (2gH)^{1/2}$. Results are
recorded with a Photron FASTCAM SA5, at a resolution of 256$\times$584
pixels ($\sim 10$ pixels per $d$), at 40,000 frames per second. To
locate and track the intruder, we use a circular
Hough transform at each frame. Velocity $v$ and acceleration $a$
are calculated by numerical differentiation, with a low-pass filter,
cutoff frequency of $133$~Hz~$\simeq (7.5$~ms$)^{-1}$~$ \simeq v_0/D$), applied
with each derivative to reduce noise amplification. The frequency cutoff
is as large as possible while maintaining a
signal-to-noise ratio of 10:1. This yields intermediate time scale
data for $v$ ($v_{int}$) and $a$ ($a_{int}$) which are still strongly
fluctuating in time. Photoelastic images are normalized by a
calibration image, taken before the intruder is dropped, to account
for inhomogeneities in the light source. After this, the discrete
gradient squared ($G^2 = |\nabla I|^2$) of the image is computed by using the spatial
variation of the image intensity $I$; the sum of the $G^2$ in a
particular region measures the local force response\cite{Howell1999} (i.e., beneath the intruder, as in
Fig.~\ref{fig:frames}). A static calibration covering
the full range of $G^2$ encountered in any impact was performed by placing a
weighted piston on a box of about 100 particles that are subject to
the same light intensity as used in the experiments. As shown in Fig.~\ref{fig:accphoto}(b), 
it is essentially linear.

{\em Comparing to previous models.\textemdash}An important question is whether
the observed dynamics are consistent with existing models,
i.e., Eq.~(\ref{eqn:forcelaw}). To address this, we consider the intruder
trajectory $z(t)$ and the filtered derivatives $v_{int}$ and
$a_{int}$. As noted, the derivatives, particularly $a_{int}$, are
strongly fluctuating, and these fluctuations are a \textit{physical} aspect of
the dynamics, as discussed below. Plots of $a_{int}$ versus $v_{int}^2$
data from different impacts with varying $v_0$ show good agreement,
within fluctuations, with Eq.~(\ref{eqn:forcelaw}). This analysis
yields $f(z)$ and $h(z)$: a constant value for $h(z)$ [i.e. $h(z) = b
\simeq 5 D$] after an initial transient at impact and $f(z)$, which
is nearly linearly increasing in depth.

However, for any individual trajectory, we measure large fluctuations
in $a_{int}$ (Fig.~\ref{fig:accphoto}), on a scale that is comparable to the mean
acceleration. These fluctuations are absent in the ``slow-time''
models discussed above, and their large amplitude is both a novel
observation and a potential weak point of the models. That is, the
braking of the intruder is not a smooth steady process but a series
of events where the intruder is subjected briefly to large
accelerations, followed by more quiescent periods that can be close to
acceleration-free.

{\em Connecting acoustic activity to intruder deceleration.\textemdash}As
noted, during an impact, we observe complex propagating force networks
(known as force chains) generated intermittently at the
leading edge of the intruder as it moves through the medium, as shown
in Fig.~\ref{fig:frames}(a), as well as in Supplemental Videos 1 and
2. To quantify the photoelastic response, we consider the angular
region extending radially outward from the bottom half of the intruder
over a length $\sim 10 d$, forming a half-annulus. Figure~\ref{fig:frames}(b) 
shows a space-time plot of the total photoelastic response in this region. 

To relate the photoelastic activity to the acceleration fluctuations,
we compare the total photoelastic response in the angular region
immediately under the intruder to $a_{int}$ (Fig.~\ref{fig:accphoto}). 
Photoelastic data are obtained at 40 kHz,
which is about 500 times faster than the frequency cutoff for
$a_{int}$. Comparing $a_{int}$ to the photoelastic response $G^2$
requires time-filtering the photoelastic data such that the time scale
matches that of $a_{int}$. This gives a comparison
at the intermediate time scale; a plot (Fig.~\ref{fig:accphoto}) of
$a_{int}$ and filtered $G^2$ data gives the same curve, showing
that the two are virtually identical. For this comparison, we first
normalized $G^2$ by a constant to obtain the optimum agreement between
filtered $G^2$ and $a_{int}$, but this normalization matches well with
the static calibration of $G^2$ discussed above. (We used this double
comparison to be sure that the static calibration matched well with
the dynamics measurements.)  We conclude that the large photoelastic
events are the main force mechanism acting on the intruder. By
inference, the energy loss for the intruder is tied to these acoustic
events rather than, e.g., to frictional drag with the intruder.

\begin{figure}
\includegraphics[width=0.75\columnwidth]{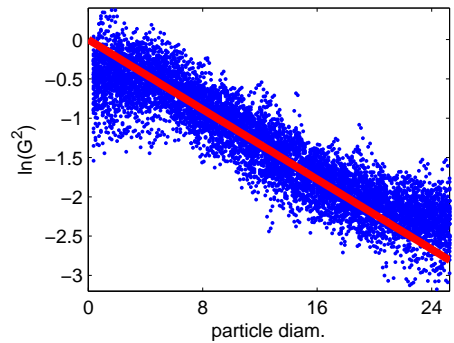}
\caption{
  (color online.) Photoelastic pulses decay as they
    propagate away from the intruder. We observe a thin angular slice of
    opening angle of $\pi/8$ rad, extending 25$d$ below, and
    centered directly beneath the intruder. We use 40 different pulses 
    from different impacts of a single intruder ($D=21.17d$), where the
    intruder velocity at the pulse emission varies between 2 and 6 m/s. We then
    plot the natural logarithm of $G^2$ per area as a function of
    depth for each pulse, normalized by the total intensity in the
    pulse (wave intensity will decrease as $1/r$ moving away from a point source in 2D, 
    and this effect has already been accounted for in this plot). The imposed fit 
    (thick, red line) is $\exp(-r/L)$, where $L$ is the
    decay length, roughly 10 particle diameters.}
\label{fig:decay}
\end{figure}

{\em Acoustic dissipation.\textemdash}Once the acoustic pulses have moved ahead
of the intruder, there must be a loss mechanism of these disturbances
within the material. Hence, it is important to examine how fast and
how far the acoustic pulses propagate. To this end, we observe the
photoelastic response in a long, thin angular slice, centered directly
beneath the intruder with a width of $\pi/8$, which extends 25$d$
beneath the intruder. Space-time plots of the response in this region
indicate a wave speed of about 325 m/s ($\sim$1/10 of the sound speed
in the bulk material from which the particles are cut). To determine
the attenuation of the acoustic pulses, we plot normalized intensity
versus depth. The normalization for each pulse is the cumulative
photoelastic response $G^2$ over its full
duration. The normalized photoelastic response averaged over multiple
events shows an exponential decay (Fig.~\ref{fig:decay}), with a decay length of $\sim$10
particle diameters, which is short enough that reflections from the
bottom or sides of the container are not important. It is unclear which grain-scale interactions
are responsible for this decay, but it could be explained by force-chain
splitting, grain-grain friction, restitutional losses for each ``collision,'' or other
dissipative mechanisms.


{\em Fluctuation statistics and stochastic description.\textemdash} Large
fluctuations in the photoelastic response (Fig.~\ref{fig:accphoto})
suggest a stochastic description, which captures mean behavior as well as 
short-time fluctuations. For example, one might modify Eq.~(\ref{eqn:forcelaw})
to
\begin{equation}
F(z,\dot{z},t)=mg-[f(z)+h(z)\dot{z}^2]\eta(t).
\label{eqn:forcelaw2}
\end{equation}
Here, $\eta(t)$ is a multiplicative stochastic term, which should
follow directly from microscopic physics and have a mean of unity. A
multiplicative term is chosen here since rescaling by the mean photoelastic behavior
yields a statistically stationary fluctuating term, as discussed below, and since fluctuations in dense
granular systems often scale with the mean (as here).

\begin{figure}
 \includegraphics[clip,trim=5mm 0mm 5mm 0mm,width=0.8\columnwidth]{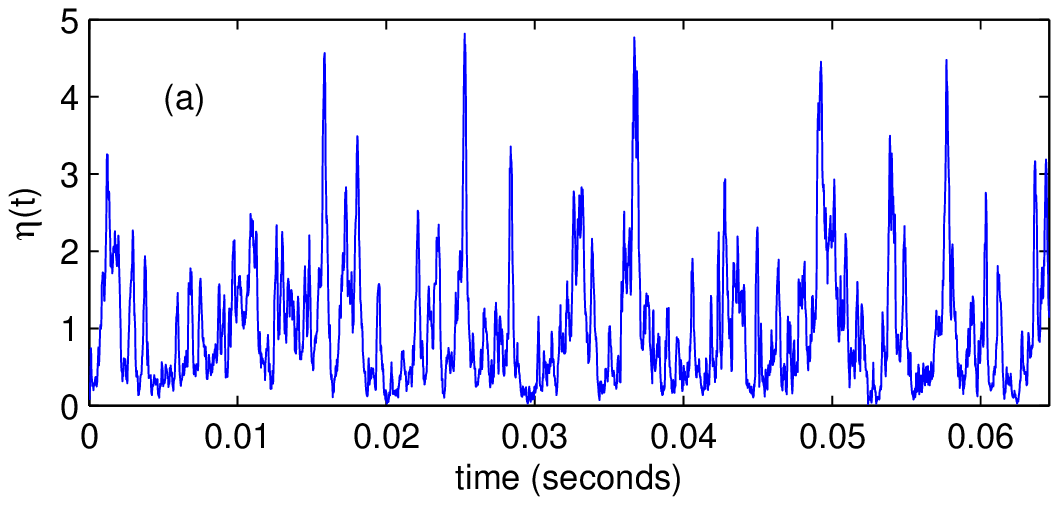}
 \includegraphics[clip,trim=5mm 0mm 5mm 0mm,width=0.8\columnwidth]{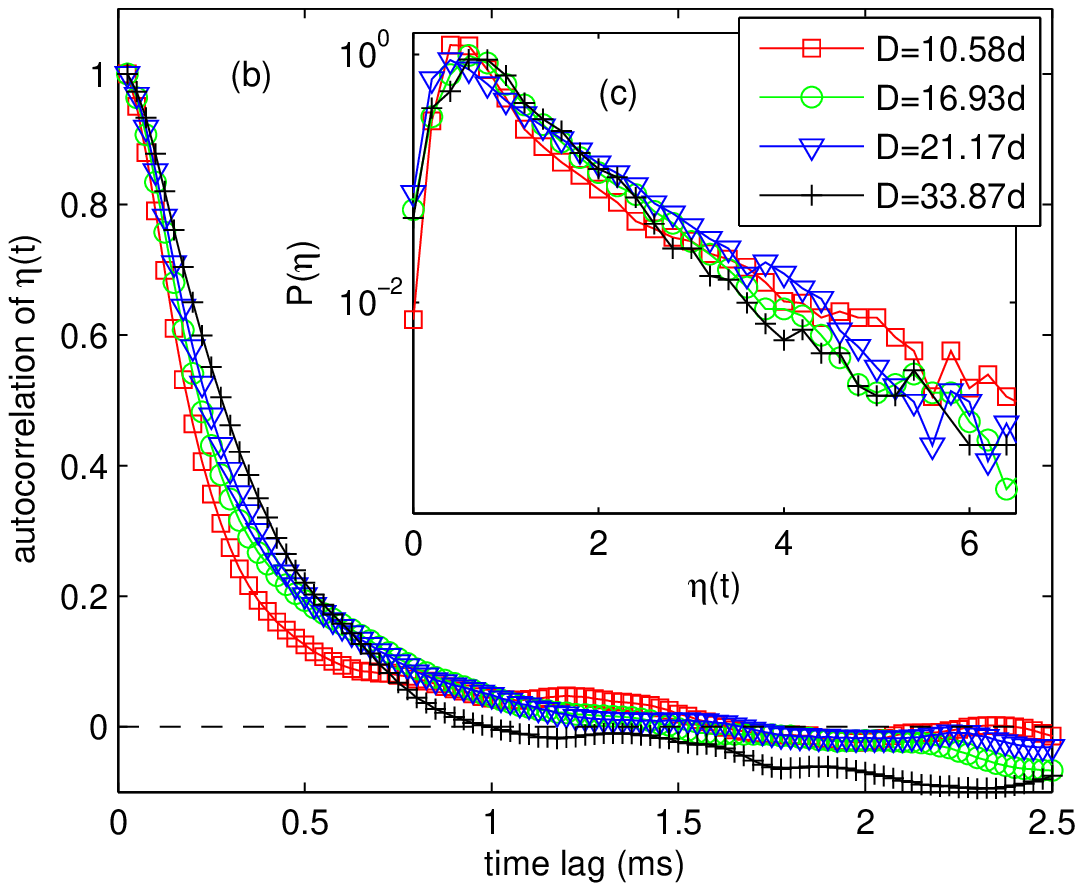}
 \caption{(color online.) (a) The fluctuating term $\eta(t)$ for a single impact
 	 (typical for all impacts), where
     $\eta(t)\sim G^2(t)/G^2_{avg}(t)$, as discussed in the text. (b) The
     autocorrelation and (c) the PDF of the
     combined fluctuating signals for all impacts for each intruder
     ($\sim$20 runs per intruder). The semilog PDF plot shows $P(\eta)\sim \exp(-\eta)$. We see an
     autocorrelation time of $\sim 1$~ms, which gives a
     typical event time, which agrees with
     video frames in Fig.~\ref{fig:frames}(a). The lack of dependence on
     intruder size suggests a collective mechanism and not a simple
     combination of uncorrelated, random force chains.}
 \label{fig:flucts}
\end{figure}

To experimentally characterize the fluctuations in
Eq.~(\ref{eqn:forcelaw2}), we write $\eta(t)\sim G^2(t)/G^2_{avg}(t)$,
where $G^2(t)$ is the photoelastic time series used to measure force
(e.g., bottom of Fig.~\ref{fig:frames}) and $G^2_{avg}(t)$ is the mean
behavior, obtained by fitting a low-order polynomial to
$G^2(t)$. This yields a fluctuating term which appears statistically
stationary throughout the duration of an impact, as shown in
Fig.~\ref{fig:flucts}. Typically, $\eta(t)$ has an autocorrelation
decay time of $\sim$1 ms and a probability distribution function
(PDF) that is nearly exponential. The PDF
describes the likelihood of the large events which dominate the
decelerating force. Such a PDF is typical for forces in static dense
granular systems and is presumably related to the probability of
generating force-chain-like structures. 

Surprisingly, the fluctuation statistics show almost no dependence
on intruder size. One might expect that the contact forces or force chains
generated from two sufficiently separated points along the bottom of the 
intruder are uncorrelated. If so, increasing the intruder size 
would include more of these independent forces, which, by the central limit theorem, would yield
smaller and more Gaussian-like fluctuations, regardless of the statistics of each one.
However, this does not occur, suggesting a more subtle collective mechanism. One possibility is that 
spatially separated intruder-particle contacts often excite the same persistent force network.

{\em Conclusion.\textemdash}In this Letter, we present a new microscopic
picture of the force on an intruder moving through a granular
material, which focuses on acoustic activity and fluctuations due to
the generation of force-chain-like pulses. We observe consistency with
established impact force models but with substantial fluctuations in
the measured deceleration of the intruder during the impact
process. We have shown that the acceleration profiles, including these
fluctuations, are a direct consequence of acoustic pulses transmitted
along networks of particles. Other recent studies have indicated an
important role for granular force networks in intruder impacts
\cite{Kondic2012} and acoustic transmission \cite{Daniels2011}. The
microscopic description presented here should also help connect granular impact experiments
with differing microstructure, such as more dilute or compacted
\cite{Goldman2010,Royer2005} or anisotropic (e.g., sheared)
systems, or even more general experiments on granular flow around an
obstacle. Strong force fluctuations suggest a stochastic model, which gives
a natural way to separate the slowly varying macroscopic response from
fast-time fluctuations. We believe that the granular sound speed is critical
in our description, so we expect substantial differences when
intruder speeds are close to sonic or even supersonic. This could be achieved by increasing
intruder velocity or reducing the granular sound speed by using softer material. Also of interest
is how these effects translate to three-dimensional systems or systems
with a much larger ratio of intruder size to particle size.

This work has been supported by the U.S. DTRA under Grant No.
HDTRA1-10-0021. We very much appreciate additional input from
Dr. C. O'Hern and Dr. W. Losert.

\bibliographystyle{prsty}

\begin{thebibliography}{10}

\bibitem{Euler1745}
L. Euler, \emph{Neue Grunds\"{a}tze der Artillerie}, reprinted in \emph{Opera Omnia}. (Druck und Verlag Von B.G. Teubner, Berlin, 1922).

\bibitem{Poncelet1829}
J. V. Poncelet, \emph{Cours de M\'{e}canique Industrielle}. (Lithographie de Clouet, Paris, 1829).

\bibitem{Allen1957}
W. A. Allen, E. B. Mayfield, and H. L. Morrison, J. Appl. Phys., \textbf{28}, 370 (1957).

\bibitem{Forrestal1992}
M. J. Forrestal and V. K. Luk, Int. J. Impact Eng. \textbf{12}, 427 (1992).

\bibitem{Tsimring2005}
L. Tsimring and D. Volfson, in \emph{Proceedings of the International Conference on Powders and Grains} (Taylor \& Francis, London, 2005), Vol 2, pp. 1215-1223.

\bibitem{Katsuragi2007}
H. Katsuragi and D. J. Durian, Nature Phys. \textbf{3}, 420 (2007).

\bibitem{Goldman2008}
D. I. Goldman and P. Umbanhowar, Phys. Rev. E \textbf{77}, 021308 (2008).

\bibitem{Goldman2010}
P. Umbanhowar and D. I. Goldman, Phys. Rev. E \textbf{82}, 010301(R) (2010).

\bibitem{Ciamarra2004}
M. P. Ciamarra, A. H. Lara, A. T. Lee, D. I. Goldman, I. Vishik, and H. L. Swinney, Phys. Rev. Lett. \textbf{92}, 194301 (2004).

\bibitem{Ambroso2005}
M. A. Ambroso, C. R. Santore, A. R. Abate, and D. J. Durian, Phys. Rev. E \textbf{71}, 051305 (2005).

\bibitem{Bruyn2004}
J. R. de Bruyn and A. M. Walsh, Can. J. of Phys. \textbf{82}, 439 (2004).

\bibitem{Walsh2003}
A. M. Walsh, K. E. Holloway, P. Habdas, and J. R. de Bruyn, Phys. Rev. Lett. \textbf{91}, 104301 (2003).

\bibitem{Nelson2008}
E. L. Nelson, H. Katsuragi, P. Mayor, and D. J. Durian, Phys. Rev. Lett. \textbf{101}, 068001 (2008).

\bibitem{Newhall2003}
K. A. Newhall and D. J. Durian, Phys. Rev. E \textbf{68}, 060301(R) (2003).

\bibitem{Hou2005}
M. Hou, Z. Peng, R. Liu, K. Lu, and C. K. Chan, Phys. Rev. E \textbf{72}, 062301 (2005)

\bibitem{Howell1999}
D. Howell, R. P. Behringer, and C. Veje, Phys. Rev. Lett. \textbf{82}, 5241 (1999).

\bibitem{Kondic2012}
L. Kondic, X. Fang, W. Losert, C. S. O'Hern, and R. P. Behringer, Phys. Rev. E \textbf{85}, 011305 (2012).

\bibitem{Daniels2011}
K. E. Daniels, and E. T. Owens, Europhys. Lett. \textbf{94}, 54005 (2011).

\bibitem{Royer2005}
J. R. Royer, E. I. Corwin, A. Flior, M. L. Cordero, M. L. Rivers, P. J. Eng, and H. M. Jaeger, Nature Phys. \textbf{1}, 164 (2005).

\end{thebibliography}

\end{document}